# Toward a fitness landscape model of firms' IT-enabled dynamic capabilities

Rogier van de Wetering[1] & Rik Bos[1]

[1] Faculty of Management, Science and Technology, Open University of the Netherlands, Valkenburgerweg 177, 6419 AT Heerlen, the Netherlands
rogier.vandewetering@ou.nl; rik.bos@ou.nl

**ABSTRACT**

*This chapter presents, extends and integrates a complexity science perspective and applies this to the firm's IT-enabled dynamic capabilities (ITDCs). By doing so, this chapter leverages statistical survey data and uses them as parameters for a simulation using Kauffman's NK-model. This NK-model creates stochastically generated fitness landscapes that are parameterized using a finite number of 'N' elements, or capabilities, and 'K' complex interactions between those capabilities, and studies the performance (fitness) of systems. We simulate a firm's effort to adaptively explore and walk through a fitness landscape of possible strategies of inter-related capabilities to reach toward higher levels of fitness of ITDCs. Also, our fitness landscape model provides realistic scenarios with a nexus of possible business strategies that can be employed considering a firm's current status, interdependency, and alignment among its capabilities. Our work suggests that firms achieve the highest fitness values when the interdependency among the individual capabilities is relatively small.*

## INTRODUCTION

Scholars posit that information systems and information technology (IS/IT) are capable of revolutionizing the way firms operate, and hence drive a firm's performance under rapidly changing conditions (Agarwal & Selen, 2009; El Sawy, Malhotra, Park, & Pavlou, 2010; Kohli & Grover, 2008; R. Van de Wetering, Mikalef, & Pateli, 2017b). In understanding this particular role of IS/IT within the turbulent business ecosystem and landscape, previous studies have emphasized the importance of IS/IT in enabling organizational capabilities (Kohli & Grover, 2008). Organizational capabilities represent firms' potential to achieve specific business strategies and management objectives using focused deployment and are considered the building blocks on which firms compete within the business ecosystem (Barney, 1991). Firms that are capable of targeting and deploying IS/IT initiatives in support of these organizational capabilities are more likely to realize business value from their IT resource and IT competencies inventory (Sambamurthy, Bharadwaj, & Grover, 2003; R. Van de Wetering, Mikalef, & Pateli, 2017a). The extant literature refers to this specific capacity to effectively use IT functionalities to support IT-related activities as an IT leveraging competence (Pavlou & El Sawy, 2010).

The core principle of examining the particular value of IS/IT in enterprise processes it is embedded, i.e., IT-enabled capabilities is also highly encouraged in contemporary IS literature (Grover & Kohli, 2012; Kohli & Grover, 2008; Mikalef, Pateli, & van de Wetering, 2016). IT-enabled capabilities gained considerable as they collectively create business value and drive a firm's competitive edge (El Sawy & Pavlou, 2008; Mithas, Tafti, Bardhan, & Goh, 2012; Roberts, Galluch, Dinger, & Grover, 2012).

The capabilities we address in this current study constitute the dimensions of IT-enabled dynamic capabilities (ITDC). We define ITDCs as a firm's ability to leverage its IT resources and IT competencies, in combination with other organizational resources and capabilities, to address rapidly changing business environments (Mikalef et al., 2016). Based on the notion that firms must be able to be stable enough to continue to deliver value in their distinctive way, and agile and adaptive enough to restructure their value proposition when circumstances demand it, there is a well-documented distinction between ordinary (operational or zero-order) and these ITDCs (Drnevich & Kriauciunas, 2011; Mikalef et al., 2016; Pavlou & El Sawy, 2011; Winter, 2003). Ordinary capabilities enable a firm to make a living in the present, while ITDCs act as a continuous driver of evolution to changing business requirements (Winter, 2003). Thus, ITDCs help firms decision-makers in practice by extending, modifying, and reconfiguring existing operational capabilities into new ones that better match the environmental conditions.

There is overwhelming evidence concerning the impact of ITDCs on firm performance at an aggregate level (Mikalef & Pateli, 2017; Rai & Tang, 2010). However, there is still little consensus on how specific interrelations (e.g., too many or too little epistatic links and interdependency among individual capabilities) result in the achievement toward higher levels of performance and evolutionary fitness. Moreover, the extant literature has predominantly focused on the aggregate construct level of ITDCs.

Consequently, there is no clear overview of what constitutes useful configurations of the underlying interdependencies. A better understanding of the nature and role of ITDCs may have profound implications for theory and practice. In doing so, this study builds on the science of complexity recently become a significant focus of interdisciplinary research, also within the IS field (Benbya & McKelvey, 2006; Merali, Papadopoulos, & Nadkarni, 2012; R. Van de Wetering & Bos, 2016; Rogier Van de Wetering, Mikalef, & Helms, 2017; Walraven, Van de Wetering, Helms, Versendaal, & Caniëls, 2018). This chapter adopts a prominent model for studies in theoretical biology and population genetics, i.e., the so-called NK-model (S. Kauffman & Levin, 1987; S. A. Kauffman, 1993), that evolved into a valuable model to understand the antecedents of efficient, flexible, innovative and IT-enabled firms (Celo, Nebus, & Wang, 2018; Onik, Fielt, & Gable, 2017; Vidgen & Wang, 2006; Yuan & Jiang, 2015).

This study aims to address the previously outlined limitations in the literature by drawing on complexity science while simultaneously building on the foundations of ITDC. In doing so, this current study makes four main contributions. First, this chapter positions the science of complexity and evolutionary (simulation) models as a suitable 'lens' to study the behavior of firms' ITDCs. This lens has also been advocated by recent studies (Kay, Leih, & Teece, 2018; Onik et al., 2017). Second, following Kauffman's work (S. Kauffman & Levin, 1987; S. A. Kauffman, 1993; S. A. Kauffman & Weinberger, 1989) this study designs the NK-model and stochastically generate fitness landscapes using the five ITDCs as core elements of the model, and 'K' complex interactions between those five, i.e., the interdependency (epistasis). Third, firms' effort is simulated to adaptively explore and walk through a generated fitness landscape of possible strategies of interrelated ITDCs to change and move toward higher levels of fitness of ITDCs. Finally, this chapter identifies which degree of epistasis on average leads to optimal fitness values and hence firms' competitive performance.

Outcomes of our work provide new empirical support for possible business strategies that can be employed considering a firm's current status and interdependency among its ITDCs[1]. To our knowledge, no scholarship is available that identifies these specific interactions among the different ITDCs and uses the NK-model to do so.

The remainder of this chapter is structured as follows. First, this chapter reviews the fitness landscapes theory and Kauffman's NK-model. Next, our methods section is outlined including the simulation model parameters, and the outcomes follow this section. This chapter ends with a discussion, including inherent limitations of this study as well as future research opportunities, and concluding remarks.

## THEORETICAL BACKGROUND

Complexity science addresses the summon of adopting a dynamic methodological approach, in which scholars and practitioners are equipped with adequate assessment tools and mechanisms for examining the processes when IT and associated capabilities add value (Rogier Van de Wetering et al., 2017). This paper draws upon foundational and empirical work and presents, extends and integrates a complex adaptive systems perspective and applies this to the firm's ITDCs.

The NK model from population genetics and theoretical biology (S. Kauffman & Levin, 1987; S. A. Kauffman, 1993; S. A. Kauffman & Weinberger, 1989) has become a particularly popular reference model for studying strategy, organizational, and innovation problems and challenges through the use of fitness landscapes (Frenken, 2006; Ganco & Agarwal, 2009; Daniel A Levinthal, 1997; McKelvey, 1999; Rivkin & Siggelkow, 2007). The model (originally) simulates an organism's adaptation and self-evolvement processes in complex and turbulent environments. Recent literature advocated the use of the NK-model to simulate and examine, e.g., organizational structure, change, and adaption (Daniel A Levinthal, 1997; Daniel A. Levinthal & Warglien, 1999; McKelvey, 1999), system and technology architecture (Frenken, 2006), business strategies (McCarthy, 2004) and enterprise intellectual capital (Fan & Lee, 2012) under the complexity science 'lens.'

The NK-model consists of $N$ elements (in our case ITDCs), and $K$ represents the richness of epistatic linkages (term used to describe interacting genes) among those elements. Alternatively, as Kauffman (S. A. Kauffman & Johnsen, 1991) defines it: "*K corresponds to the number of other genes, or traits, which have a bearing on the fitness contribution of each gene or trait.*" This richness of epistatic interactions among sites makes it possible for, e.g., a population to evolve toward different combinations of alleles depending on its initial genetic composition (S. A. Kauffman & Weinberger, 1989). The possible alleles of a particular element of a system can formally be described. Each string $x$ is described by $x_1 x_2 x_3 \ldots x_n$. These strings are part of a solution or possibility set $X$, so that for two allelic values, or a two-allele additive haploid genetic model, as Kauffman (S. A. Kauffman & Johnsen, 1991) favors it we have:

$$x = x_1 x_2 x_3 \ldots x_N \in X; x_i \in \{0,1\}$$

The $N$-dimensional possibility space $X$ is called the design space of a system. It, therefore, includes all possible combinations of the alleles of the individual system elements (Frenken, 2001). If we, thus, assume that each element within a system has two possible values, the binary strings with $N = 3$ can take the form of 000, 010, 101…..111. These strings are part of the entire possibility space which we formalize as $X = 2^N$.

In general, the degree of interaction and interconnectedness between those $N$ elements, i.e., $K$, ranges from $K = 0$ (the least complex) to $K = N - 1$ (having the most complex architecture), and influences the ruggedness of a fitness landscape. With $K = 0$ the system's elements are independent concerning the fitness of all other elements. When a system is fully interconnected, each element is connected to all other elements. With low levels of epistatic linkages among elements (e.g., $K = 0$) the fitness landscape tends to be smooth with a single performance or fitness peak.

The fitness contribution of each element or trait depends upon itself and thus epistatically on K other traits (Altenberg, 1997; S. Kauffman & Levin, 1987; S. A. Kauffman & Weinberger, 1989). Therefore, each particular element makes a fitness contribution which, in turn, also (partly) depend upon the specified combinations, among $2^{(K+1)}$ of the presence or absence of the $K+1$ traits which bear upon its fitness. Now, these fitness contributions have to be specified. This procedure is done through the implementation of a matrix of all possible combinations in the *N*-dimensional binary space and their corresponding fitness (Bocanet & Ponsiglione, 2012; S. A. Kauffman, 1993). So, the fitness landscape is then created by randomly assigning values to the $2^{(K+1)}*N$ fitness values. Having assigned all the fitness values, we can define the fitness of an entire genotype and specified string of possible $x_i$ within $X$. Kauffman's NK-model specifies the overall fitness value $F$ of a string $x$ with $N$ loci as the mean of the fitness values $f_i$ of each element with two (binary) possible alleles at each position $i$:

$$F(x) = \frac{1}{N} \sum_{i=1}^{N} f_i(x_i, x_{i_1}, x_{i_2}, \dots, x_{i_K})$$

---

[1] For this research we are satisfied with assuming that firm performance is positively influenced by the level of ITDCs fitness.

Where $\{i_1,\dots, i_K\} \subseteq \{1,\dots, i-1, i+1,\dots N\}$. $x_i$ now is a value of bit i within solution x. Following this logic, fitness values fi drawn from a uniform distribution ranging from 0 to 1 (S. Kauffman & Levin, 1987; S. A. Kauffman & Weinberger, 1989), thus, take on a different value when a particular element i changes or when another element changes and thereby epistatically affects element i (Altenberg, 1997).

Following the logic from Kauffman's NK-model, a strategic change is assumed to be a process of moving from one firm strategy to another in search of an improved fitness (Daniel A Levinthal, 1997; Daniel A. Levinthal & Warglien, 1999; McCarthy, 2004). An 'adaptive walk' performs this search (S. A. Kauffman, 1993). In essence, what this means, is that if you would (randomly) initially select a point in the landscape, e.g., 001, there are now in total three 1-mutant neighbors, thus directly connected to this particular point in the landscape design (S. A. Kauffman, 1993).

**METHODS**

*Simulation model and NK-model parameters*

All included dimensions that comprise ITDCs are based on past empirical and validated work. Hence, this chapter adopts five elementary dimensions of ITDCs, i.e.: (1) sensing (SEN), (2) learning (LRN), (3) coordinating (CRD), (4) integrating (INT), and (5) reconfiguring (REC) (Mikalef & Pateli, 2017; Mikalef et al., 2016; Pavlou & El Sawy, 2011; Protogerou, Caloghirou, & Lioukas, 2012) with each capability having two allelic values, i.e., '0' or '1'. The respective weights of the ITDCs in their contribution to fitness have to be specified. In the original NK-model, Kauffman assumes that all elements of the system weigh equally in their fitness contribution. In practice, however, this does not necessarily have to be the case as specific levels of complementary among the capabilities may be present under different circumstances. This view is also empirically supported by (Fan & Lee, 2012; Valente, 2008). In light of this shortcoming of the original NK-model, the model parameters are extended with 'unequal' known weighting factors ($\varphi$) based on validated empirical work that reports path or regression coefficients of the individual capabilities (Mikalef & Pateli, 2016, 2017), see Table 1.

Table 1.   Path coefficients of the weighted model; *** $p < 0.001$

| IT-enabled dynamic capabilities | SEN | LRN | CRD | INT | REC |
| --- | --- | --- | --- | --- | --- |
| *Path weights ($\beta_i$)* | 0.226*** | 0.249*** | 0.212*** | 0.212*** | 0.245*** |

As suggested by (Fan & Lee, 2012), one can now assign the different weighting factors ($\varphi_1,\dots \varphi_5$), as $\varphi_i$ being equal to ß$_i$ divided by the sum of all path weights. Note that the $\varphi_i$ are appropriately weighted relatively to all the other capabilities, although the weights do not differ significantly. The overall fitness formula is adjusted accordingly (note that the factor $1/N$ now is absent because the $\varphi_i$ by definition already sum up to 1):

$$F(x) = \sum_{i=1}^{N} \varphi_i f_i(x_i, x_{i_1}, x_{i_2}, \dots, x_{i_K})$$

As a final step, random fitness values, $f_i$ within the NK-model need to be specified. $f_i$ is dependent on $K + 1$ binary variables, so there are $2^{(K+1)}$ different inputs. This study determines the corresponding outputs by assigning a value from a random variable that is uniformly distributed on [0,1]. For this, the RAND() function is used in Microsoft Excel 2013, and we represent $f_i$ in table form. Now, that this study includes five ITDCs, $2^{(K+1)}*N$ ($N = 5$) table entries have to be generated. This set-up works for all integer *K*-values and thus for the simulation runs for $K = 0,\dots,4$ in this study. Once the fitness landscapes have been specified, they remain fixed.

*Proposition development*

Firm achieve fitness gains through varying levels of interdependency among the different ITDCs. Synergies only arise among the capabilities when they exhibit a degree of mutual interdependence and thus not when they are completely independent (i.e., $K = 0$). In the case of a fully connected model, the associated fitness landscape is entirely random (S. A. Kauffman, 1993; Weinberger, 1991). Also, the expected fraction of fitter neighbors falls by half at each adaptive step (S. A. Kauffman & Weinberger, 1989). Thus, when the level of $K$ increases, it seems that more and more conflicting constraints emerge and frustration sets in (S. A. Kauffman & Weinberger, 1989).

Research has shown that for intermediate levels of $K$, the fitness landscape usually contains several local optima, but less than in the case of a maximum complexity scenario, i.e., $K = N - 1$ (Frenken, Marengo, & Valente, 1999). Kauffman (S. A. Kauffman, 1993) in general showed that fitness values of local optima of low-K systems are highest on average. This outcome leads us to believe that the interdependencies between ITDCs are the strongest (and lead to the highest fitness values) if the number of epistatic relations is relatively small. Another striking insight from Kauffman's simulations is that in the case of $K = 2$ (with differing values of N), the highest optima on the fitness landscape are nearest to each other and on average obtained fitness values are the highest (S. A. Kauffman, 1993). Low $K$-valued configurations (or systems), therefore, enjoy at least two evolutionary advantages over systems with high K-values (Frenken et al., 1999; S. A. Kauffman, 1993). First, local optima obtain, on average, higher fitness values. Second, and even more important, is the fact that the (correlated) structure of their landscape and the large basins of attraction of good local optima make firms more likely attain high fitness levels in lower K-systems (Frenken et al., 1999). Therefore, a low K-value that epistatically connects ITDCs is expected to outperform higher K-value configurations. Hence, we postulate:

**Proposition**: *Too many ($K = N - 1$) and too little ($K = 0$) epistatic links and interdependency among the ITDCs will inhibit overall fitness in comparison to a simple, non-complex level of epistasis; In other words, a configuration of $K = 2$, on average, will outperform all other K-values in terms of fitness values.*

## ANALYSES AND RESULTS OF NK-MODEL SIMULATIONS

*Simulation analyses*

In comparison with other NK-studies in the field of management and IS, this study ran our model using various values for the specified parameters. While keeping $N$ constant because of our construct, i.e., ITDCs contains five underlying capabilities, this study varied with values for $K$.
As our model is probabilistic and dynamic, many runs are needed to ensure significant results (Ganco & Agarwal, 2009). Therefore, we performed 5 runs of 10,000 simulations, i.e., 50,000 simulations, for each value of K, i.e., $K = 0$, $K = 1$, $K = 2$, $K = 3$ and $K = 4$ to average out random effects (Ghemawat & Levinthal, 2000). The generated data points in the fitness landscapes need to be searched and analyzed to test our proposition. Adaptive walks vary dramatically as the ruggedness of the landscape varies (S. A. Kauffman & Weinberger, 1989).

This study employs a local search strategy[2] (Nickerson & Zenger, 2004) that aligns well with one's intuition of what in practice such a search process would look like. The search strategy explores the associated neighborhood which we define as those forms or IT-enabled dynamic capability combinations that vary from the firms' current one by only one attribute (Daniel A Levinthal, 1997). Hence, considering now that $N$ capabilities have been included, a firm now has $N$ different solutions in its immediate neighborhood. When, by randomly picking one of these, a higher fitness level is encountered, the search comes to an end, taking for granted that, on average, a modest performance level might be acquired (Daniel A Levinthal, 1997). This limitation can be addressed by incorporating 'long jumps,' i.e., distant search efforts on the fitness landscape, although, this strategy does not adequately exploit past obtained experience. Therefore, this search strategy is likely to result in a deterioration of performance (Daniel A Levinthal, 1997).

*Core results*

The main findings from our extensive simulation runs are now reported. Figure 1 shows the average fitness values for the 50,000 runs per K-value. Hence, fitness values range from 0.67 (for $K = 0$) toward 0.70 (for $K = 2$). An optimum can be seen at the predicted $K = 2$ level. This outcome confirms our proposition and supports the claim that, on average, an ITDC configuration of $K = 2$ will outperform all other K-values regarding fitness. This level of epistasis benefits firms in the process of obtaining higher fitness values and, of course, in such a configuration, the adaptive walk does not get caught in a local optimum.

The lowest obtained fitness values can be seen when looking at the extreme cases, i.e., $K = 0$ and $K = N - 1$. This particular outcome indeed suggests that having too many or too little epistatic links and interdependency among the capabilities will inhibit overall fitness in comparison to a more simple level of epistasis. Clearly, simulation outcomes show that higher fitness levels are obtained with the fully connected model, i.e., $K = N - 1$ instead of the $K = 0$ model. This results could be the effect of having a low $N$-parameter.

---

[2] Obviously, more advanced and complex search strategies are possible, e.g., long-jumps, directional search and other forms of adaptive walks.

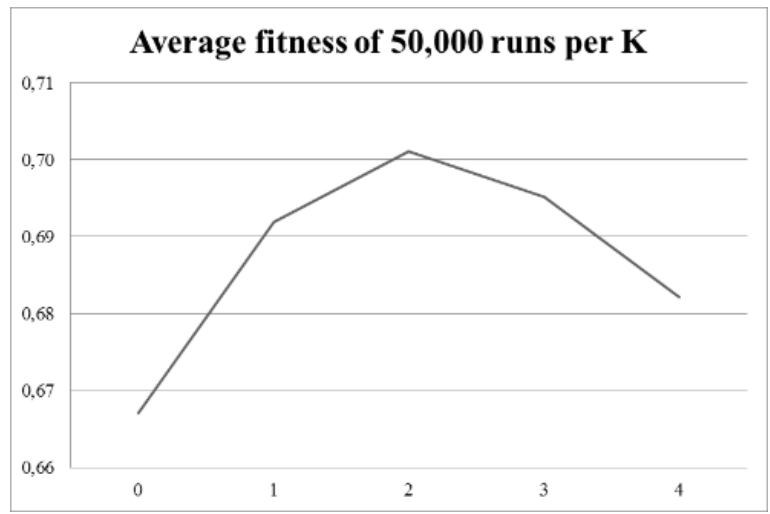

*Figure 1. Average fitness landscape values for different K-values*

**DISCUSSION**

Most studies on ITDCs and the process of leveraging IT resources and competencies within firms, do not explicitly state their assumptions concerning the interdependency among the studied capabilities. This development, without a doubt, has contributed to seeming contradictions in IS and management scholarships. From the extant literature, we know that interdependent firm activities are critical sources of performance enhancements. This current research, however, shows that these levels and the degree of mutually reinforcing relationship between ITDCs are conditioned.

This study used a finite number of '*N*' elements, thus the five ITDCs (sensing, learning, coordinating, integrating, and reconfiguring), and a degree of '*K*' complex interactions between those capabilities. The science of complexity was then introduced as a fitting 'lens,' and this chapter highlighted the role the NK-model to study evolution on fitness landscapes on which actors or agents can tune their fitness base on available options and configurations. We departed our research with the knowledge that the availability of new alternatives among ITDCs reflects a process of (re)combining multiple choices available. This chapter, therefore, extended this fundamental insight and simulated the degree of interconnectedness (K-value) among ITDCs to achieve optimal fitness and drive toward higher levels of competitive firm performance. Moreover, our fitness landscape model provides realistic scenarios with a nexus of possible strategies that can be employed considering a firm's current ITDCs status and interdependency among its capabilities.

Our findings have implications for business and IT managers. First, our simulations outcomes showed that having a non-complex level of epistasis among the capabilities, i.e., $K = 2$ is a crucial source of value through strategic options. Like (Celo et al., 2018; Ganco & Agarwal, 2009), this study highlights that certain levels of epistasis among ITDCs benefits firms considerably in a continuous process of obtaining higher fitness values and enhance firms' competitiveness. Second, the outcomes demonstrate the importance of investing in firms' ability to leverage its IT resources and IT competencies while simultaneously taking into account the diversity of processes and procedures that underlie the ITDCs. Hence, it is imperative for managers to pursue a pluralistic ITDC investment and development strategy. This outcome relates well to the claim made by Ghemawat and Levinthal (Ghemawat & Levinthal, 2000) that 'some choices condition other choices.' This study, thus, guides managers on allocating their firm resources across ITDCs and synchronize them with the degree of internal complexity and environmental turbulence. In doing so, decision-makers should focus capitalizing on substantial IT investments by focusing on maturing and enhancing core areas of the firm, as denoted by the underlying dimensions of ITDCs. This particular aspect raises the need for IT and business decision-makers to form multi-disciplinary teams including IT and business experts. When executed harmoniously, IT and business representatives are committed and engaged in the process of sensing, seizing and reconfiguring internal and external IT and business competences and capabilities to address the rapidly changing business ecosystem. Third, the synthesis from this current study implies that next to a focus on one specific ITDC—where firms allocate their investments, resources, and assets to, depending on the specific conditions and state of the capability they wish to progress—firms also need to take into account the conditions and enhancement opportunities for two interrelated yet distinct ITDCs. In doing so, firms should follow a process-oriented perspective and focus on those particular activities that are closely related in the value chain. For instance, when a firm lacks competences and capabilities in effectively sensing the environment to spot, interpret, and pursue opportunities and threats, it only makes sense that resources are directed toward revamping existing operational capabilities with new knowledge (i.e., learning capability) and subsequently embed this new knowledge into operational capabilities (i.e., the integrating ITDC). In case a firm experiences difficulties in the deployment of resources, and activities in reconfigured operational capabilities, we suggest that firms first need to focus on synchronizing their assets, resources, and activities (i.e., integrating capability) next to maturing their coordinating ITDC. Also, firms need to direct their coordinating activities to create value to calibrate the requirements for change and to implement necessary business adjustments. There are many other configuration and patterns possible as there are reciprocal relationships among the distinct ITDCs (Pavlou & El Sawy, 2011).

**FUTURE RESEARCH DIRECTIONS**

Despite this study's contributions, there are several limitations that future work must seek to address. First, this study currently only applied a fixed structure model; that is, environmental factors were not included as well as the influence of collaborative partners/firms modern firms typically have. Current-day organizations need to be able to respond to operational and market adjustments in a swift manner (Tallon & Pinsonneault, 2011) and take external environmental factors into account within corporate IS/IT strategic planning (Newkirk & Lederer, 2006). It could, therefore, be possible that these factors influence the model's outcomes. Second, our study currently did not extensively discuss the trade-off between short and possible longer adaptive walks leading to specific fitness values. This aspect could have significant implications for business strategies.

Third, an acknowledged limitation of the NK-model (S. Kauffman & Levin, 1987; S. A. Kauffman, 1993), however, is that it uses a binary measure for the interdependency parameter between two components, i.e., $K$ (Fan & Lee, 2012; Li et al., 2006; Valente, 2008). So, an interaction between capabilities is either present or absent. While, in practice, one could argue that interaction between organizational components or entities, and thus also ITDCs, can be reflected by a specific degree of interaction, and is based on the characteristic of the particular organization under study (Fan & Lee, 2012). Therefore, an argument can be made for the additional inclusion of a simulation with a continuous and fractional value in [0,1] as a measure for $K$.

Finally, our study included only a limited amount of elements '$N$' into our simulation model. In comparison to other NK-model investigations (S. A. Kauffman, 1993; Daniel A. Levinthal & Warglien, 1999; Rivkin & Siggelkow, 2007; Siggelkow & Levinthal, 2003), this amount of interconnected elements is rather small. Avenues for future research could expand the scope and reach of ITDCs and related (operational or zero-order capabilities) and inter-organizational collaboration capabilities and help identify a comprehensive set of business strategies based on the simulation outcomes.

**CONCLUSION**

Our study is the first to present a fitness landscape model of firms' ITDCs. In this process, statistical survey data were leveraged and used as parameters for a simulation using an extended NK-model. Our contribution, therefore, differs from prior scholarships that use common statistical tools and test the direct or indirect, but isolated impact of ITDCs on competitive performance (Mikalef et al., 2016; Pavlou & El Sawy, 2006, 2011). These results of this study offer several contributions that improve our theoretical understanding of the role of ITDCs and their role in the process of driving toward superior firm performance and extends conceptual and variance based studies on organizational capabilities and ITDCs in particular (McCarthy, 2004; Mikalef et al., 2016). This chapter also demonstrates that the NK-model can indeed be applied toward more general representation problems, as suggested by Altenberg (Altenberg, 1997).